\documentstyle[preprint,aps]{revtex}
\begin{document}
\preprint{gr-qc/9610050 }
\title{ Back-reaction of a conformal field on a three-dimensional black hole }
\addtocounter{footnote}{1}
\author {
Cristi\'an Mart\'{\i}nez\thanks{Electronic address:
martinez@cecs.cl}}
\address{
Centro de Estudios
Cient\'{\i}ficos de Santiago, Casilla 16443, Santiago 9, Chile.}
\author {
Jorge Zanelli\thanks{Electronic address: jz@cecs.cl}} 
\address{
Centro de Estudios
Cient\'{\i}ficos de Santiago, Casilla 16443, Santiago 9, Chile \\and
Departamento de F\'{\i}sica, Universidad de Santiago de Chile,
Casilla 307, Santiago 2, Chile.} 
\maketitle
\begin{abstract}
The first order corrections to the geometry of the (2+1)-dimensional black hole due to back-reaction of a massless conformal scalar field are computed. The renormalized stress energy tensor used as the source of Einstein equations is computed with the Green function for the black-hole background with transparent boundary conditions. This tensor has the same functional form as the one found in the nonperturbative case which can be exactly solved. Thus, a static, circularly symmetric and asymptotically anti-de Sitter black hole
solution of the semiclassical equations is found. The corrections to the thermodynamic quantities are also computed.

\end{abstract}


\section{Introduction}  \label{introd}

Black holes have fascinating properties. One of them is the fact that they radiate with thermal spectrum and might be expected to be in thermal equilibrium with the surrounding radiation in the form of a heat bath of quantum fields \cite{hawking}. 

The presence of the radiation fields is usually included as a small perturbation that can be ignored in the calculation of the geometry at lowest order. It is possible to consider the gravitational effects of the quantum fields by setting the expectation value of the corresponding stress-energy tensor as the source of Einstein's equations. It is expected that the inclusion of these semiclassical corrections would produce an improved description of the spacetime geometry. The  calculation of these corrections due to the back-reaction of matter on the spacetime of a black hole in 2+1 dimensions is the aim of this paper.

In four dimensions the back-reaction problem has been vastly
studied \cite{birrell,york}. Nevertheless, the calculations are quite difficult and necessarily require approximations whose validity could be questioned. If the aim is to compute the corrections to order $\hbar$ to the geometry, one should also include the corrections of the same order coming from quantum gravity. In the absence of a consistent quantum theory of gravitation however, the semiclassical approximation is doubtful.  

The semiclassical approach is considerably better posed in three dimensions, where a consistent quantum theory of gravity is expected to exist \cite{witten}. Three-dimensional gravity has no propagating degrees of freedom and at each point the Riemann tensor is completely determined by the matter sources there. Nevertheless, 2+1 gravity with negative cosmological constant has a rich mathematical structure and a number of interesting classical solutions including the recently found black holes \cite{BTZ,BHTZ} (BTZ solution). The possibility of constructing a quantum theory of gravity and the
existence of nontrivial solutions has made of 2+1 gravity an interesting test ground for understanding the main problems of the 3+1 theory in a simplified setting \cite{carlip1,carlip2}.

The back reaction on the 2+1 black hole due to conformally coupled fields in certain approximations was studied by Lifschytz and Ortiz \cite{LO}, and by Shiraishi and Maki \cite{SM1}. There, asymptotic expressions of the metric for $r \rightarrow \infty$ and for large $M$ are given. The ansatz proposed in \cite{LO} does not satisfy the modified Einstein equations, making its interpretation somewhat
obscure. In both these articles the metric --in spherical coordinates-- takes the form $-g_{tt} = g^{rr} \sim r^2 -M +O(r^{-1})$ and this is also the case for  the exact solution of the semiclassical equations analyzed below.

In a previous article we constructed an exact black-hole solution in 2+1 dimensions interacting with a conformally coupled scalar field \cite{MZ}. The system is described by the action $I= I_{\rm grav} + I_{\rm matter}$, where 
\begin{equation} \label{g-action}
I_{\rm grav}= \frac{1}{2 \kappa}\int d^3 x \sqrt{-g}[R+2l^{-2}]
\end{equation}
and
\begin{equation} \label{m-action}
I_{\rm matter} = -\int d^3 x \sqrt{-g} \left[ 
\frac{1}{2}g^{\mu \nu}\nabla_{\mu}\Psi\nabla_{\nu}\Psi
+\frac{1}{16}R\,\Psi^2 \right] \,,
\end{equation}
where $\kappa = 8\pi G$ and $-l^{-2}$ is the cosmological constant.

The conformal coupling of the scalar field produces a remarkable simplification of the coupled equations that makes it possible to solve them exactly. The exact solution constructed in Ref. \cite{MZ} is nonperturbative and the scalar field cannot be continuously switched off stripping the black hole of its surrounding matter field.  

In this note we turn our attention to the back reaction problem, on the geometry due to a small perturbation around the matter-free BTZ black hole. The renormalized one-loop effective stress-energy tensor is given in terms of the propagator for the scalar field conformally coupled to the black hole background. Avis, Isham and Storey \cite{AIS} calculated the two point function for anti-de Sitter space in 3+1 dimensions with various asymptotic boundary conditions. Their approach was also used in \cite{LO,SM1,steif} to calculate the propagation of scalar fields in the background of the 2+1 black hole. The propagator on the BTZ background can be obtained from the one on the universal covering of anti-de Sitter space ($\overline{AdS}$) using the fact that the black
hole geometry is the quotient $\overline{AdS}/\{H\}$, where $\{H\}$ is a discrete subgroup of SO(2,2) \cite{BHTZ}. 

The paper is organized as follows. The construction of the Green function and the evaluation of the one-loop stress tensor is summarized in section II. In section III we evaluate the corrections to the geometry and thermodynamics of the black hole.  The results are discussed in section IV.

\section{Quantum field theory on the BTZ background}

In order to evaluate the back reaction we need to compute the stress tensor for a conformal scalar field (\ref{m-action}) which in the black hole background reads
\begin{equation} \label{t21m}
T_{\mu\nu}=\frac{3}{4}\nabla_{\mu}\Psi\nabla_{\nu}\Psi
-\frac{1}{4} \Psi \nabla_{\mu}\nabla_{\nu}\Psi -\frac{1}{4}g_{\mu\nu} 
\left[ (\nabla \Psi)^2 +\frac{1}{4 l^2}\Psi^2\right] \,,
\end{equation}
where we have used that $G_{\mu \nu}=l^{-2} \, g_{\mu\nu}$ for the  
solution and the field equation $ \Box \, \Psi -\frac{R}{8} \Psi=0$. 

The expectation value $\left\langle T_{\mu \nu} \right\rangle$ is obtained by differentiating the two-point function $G(x,x')$ and taking the limit $x' \rightarrow x$ \cite{birrell}. This point-splitting regularization yields
\begin{equation} \label{tq}
\left\langle T_{\mu \nu} (x) \right\rangle= \lim_{x' \rightarrow x} \, \frac{\hbar}{4} \left[3 \nabla_{\mu}^{x}\nabla_{\nu}^{x'} -g_{\mu\nu}g^{\alpha \beta} \nabla_{\alpha}^{x}\nabla_{\beta}^{x'}-\nabla_{\mu}^{x}\nabla_{\nu}^{x}\\  -\frac{1}{4 l^2} g_{\mu \nu}\right]\, G(x,x') \,.
\end{equation}

The computation of  the two-point function $G(x,x') = \left \langle 0 \right|\Psi(x) \Psi(x')\left| 0 \right\rangle $ on the 2+1 black-hole background is greatly simplified by  the fact that this solution can
be obtained from $\overline{AdS}$ through identifications by means of a discrete subgroup of the SO(2,2) isometry group \cite{BHTZ}. We briefly review this construction. 

The universal covering of anti-de Sitter space can be defined as the
3-dimensional pseudosphere embedded in {\hbox{{\rm I} \kern-.2em\hbox{\rm R}}}$^{2,2}$, ${\cal S} = \{ u\in ${\hbox{{\rm I}\kern-.2em\hbox{\rm R}}}$^{2,2} / -u_{1}^2+u_{2}^2-u_{3}^2 + u_{4}^2 =-l^2\}$ . The hypersurface $\cal{S}$ can be parametrized by coordinates $0\leq r < \infty$, $-\infty < \theta < \infty$ and $-\infty < t < \infty$ in the form

\begin{center}$
\begin{array}{lll}
u_1&=&\displaystyle\pm\frac{r}{\sqrt{M}} \cosh \sqrt{M}\theta \\
u_2&=&\displaystyle\frac{r}{\sqrt{M}}
\sinh\sqrt{M}\theta \\
u_3&=&\displaystyle\sqrt{\frac{|r^2-Ml^2|}{M}} \sinh \sqrt{Ml^{-2}}t \\
u_4&=&\displaystyle\pm\sqrt{\frac{|r^2-Ml^2|}{M}} \cosh \sqrt{Ml^{-2}}t \,.
\end{array}$
\end{center}

In these coordinates, the metric on $\cal{S}$ is that of the black hole solution
\begin{equation} \label{BTZ}
ds^2= -\left(\frac{r^2}{l^{2}}-M\right) dt^2 + 
   \left(\frac{r^2}{l^{2}}-M\right)^{-1} dr^2 + r^2 d\theta^2   \,,
\end{equation}
where we identify the mass $M$ and event horizon, $r_+= \sqrt{M} l$. There is one important difference however between the geometry of  $\cal{S}$ and that of the BTZ solution because in the latter $\theta$ has to be identified with $\theta +2\pi$. In the coordinates of the embedding space {\hbox{{\rm I}\kern-.2em\hbox{\rm R}}}$^{2,2}$ the identification means taking the quotient by the element of SO(2,2) acting on $\cal{S}$ by boost on the $u_1-u_2$ sections,
$$
H=\left[ \begin{array}{cccc}
\cosh 2\pi  \sqrt{M}&\sinh 2\pi  \sqrt{M}&0&0\\
\sinh 2\pi  \sqrt{M}&\cosh 2\pi  \sqrt{M}&0&0\\
0&0&1&0\\
0&0&0&1 \end{array}\right] \, .
$$

Since the group of isometric translations on $\overline{AdS}$ generated by $H$ is Abelian, its action on a field can be represented by
$$
H:\Psi(x) \mapsto \Psi'(x')= e^{-\imath \delta} \Psi(x),
$$
where $\delta$ is an arbitrary phase.  For a field with periodic boundary conditions $\delta=0$ (``bosons": $\Psi(\theta) = \Psi(
\theta + 2\pi)$), while for fields with antiperiodic boundary conditions $\delta = \pi$ (``fermions": $\Psi(\theta) =-\Psi(\theta +2\pi)$). Contrary to the situation in 3+1 dimensions, the phase can  take any value because a rotation by $2\pi$ is not homotopic to the identity and anyons might exist.  Then, the propagator on the black-hole background is a superposition of the amplitudes of propagation from one point to another along trajectories on every possible homotopy class, that is,
\begin{equation}  \label{imag}
G_{\rm BTZ}(x,x')= \sum_{n\in {\bf Z}} e^{-\imath n \delta} \, G_{\rm \overline{AdS}}(x, H^{n}x')\,,
\end{equation}
where $H x'$ denotes the action of $H$ on $x'$. Thus the problem of finding the propagator on the black hole background reduces to a  simpler one on $\overline{AdS}$ space.

Inspite of the simplification mentioned above, a difficulty arises from the fact that $\overline{AdS}$ is not a globally hyperbolic spacetime and it is necessary to impose boundary conditions at (time-like) spatial infinity in order to formulate a sensible quantum field theory. This problem was carefully analized, in 3+1 dimensions,  in Ref. \cite{AIS}, where it was shown that there are three resonable boundary conditions for a scalar field at the spatial infinity: Dirichlet(D), Neumann(N), and ``transparent"(T). The same boundary conditions can be imposed in 2+1 dimensions \cite{LO}. In particular,
for a conformal scalar field described by the action (\ref{m-action}), the corresponding Green functions are \cite{carlip2}  
\begin{eqnarray}
G_{\rm \overline{AdS}}^{\rm D}(x,x')&=& \frac{1}{4\pi}\left[ \sigma^{-1/2}-
                     (\sigma+4l^2)^{-1/2}\right] \,, \\
G_{\rm \overline{AdS}}^{\rm N}(x,x')&=& \frac{1}{4\pi}\left[ \sigma^{-1/2}+
                     (\sigma+4l^2)^{-1/2}\right] \,, \\
G_{\rm \overline{AdS}}^{\rm T}(x,x')&=& \frac{1}{4\pi}\,\sigma^{-1/2} \,.
\end{eqnarray}
Here $\sigma(x,x')$ is the square of geodesic distance between $x$ and $x'$ in the embedding space of $\cal{S}$, ${\hbox{{\rm I}\kern-.2em\hbox{\rm R}}}^{(2,2)}$. Clearly, only two of these Green functions are independent and it is also possible to define a Green function with ``mixed" boundary conditions, namely 
\begin{equation} \label{mixed}
G_{\rm \overline{AdS}}^{\alpha}(x,x')= \frac{1}{4\pi}\left[ \sigma^{-1/2}-\alpha(\sigma+4l^2)^{-1/2}\right] \,,
\end{equation}
which includes the previous cases \cite{SM1}.

For $x'\rightarrow x$ the $n=0$ term in the sum (\ref{imag}) diverges and therefore $\left\langle T_{\mu \nu} \right\rangle$ must be renormalized subtracting this term. This  expectation value was evaluated, for the spinless black hole under Neumann and Dirichlet boundary conditions in \cite{LO}; under mixed boundary conditions in \cite{SM1}; and for a rotating black hole with transparent boundary conditions in \cite{steif}.

For transparent boundary conditions, (\ref{tq}) is
\begin{eqnarray} \label{tqt}
\kappa \left\langle T_{\mu }^{\nu}  \right\rangle &=& \frac{l_P}{r^3} \, F(M)
\,  \mbox{diag}(1,1,-2) \,,\quad \mbox{with}\\
F(M)&= & \frac{M^{3/2}}{2 \sqrt{2}} \sum_{n=1}^{\infty} \,e^{-\imath n \delta}
\frac{\cosh 2\pi n\sqrt{M} +3}{(\cosh 2\pi n\sqrt{M}-1 )^{3/2}}\label{tqtp}\,,
\end{eqnarray}
where the right-hand side of eq. (\ref{tqt})is evaluated in units such that $G=1/8$,  $M$ is the ADM black-hole mass, and $l_P=\hbar/8$ is the Planck length in three dimensions\footnote{In three spacetime dimensions, the Planck mass, defined by $m_P l_P = \hbar$, is {\em independent} of  $\hbar$, $m_P=8$ and $\kappa=\pi$.}.  

The series (12) converges exponentially for any $M>0$ and the stress tensor is finite everywhere except for $r \rightarrow 0$. The  divergence at the origin arises from the fact that  $r=0$ is a fixed point under the action of $H$. For $M>>1$, the first term dominates the series and $F(M) \sim e^{-\pi\sqrt{M}}$, that is, $\left\langle T_{\mu \nu} \right\rangle $ vanishes exponentially for large $M$. For small $M$ $\left\langle T_{\mu \nu} \right\rangle$ is finite. 

We observe that $\left\langle T_{\mu \nu} \right\rangle$ is conserved and traceless. This is consistent with the known fact that there are no anomalies in odd dimensions \cite{birrell}.

\section{Back reaction} 

The $O(\hbar)$-corrections to the black hole geometry due to the surrounding conformal scalar radiation field are given by the semiclassical equations 
\begin{equation} \label{semi}
G_{\mu \nu} - l^{-2} g_{\mu \nu}= \kappa \left\langle T_{\mu 
 \nu} \right\rangle \,,
\end{equation}
with $\left\langle T_{\mu \nu} \right\rangle$ given by (\ref{tqt}), (\ref{tqtp}).

\subsection{Geometry}

Let us look for a static and circularly symmetric solution. The ansatz
$$
ds^2= -A(r) dt^2 + B(r) dr^2 + r^2 d \theta^2 \,
$$
is a solution of the semiclassical Einstein equations (\ref{semi}) if 
\begin{eqnarray}
-\frac{B'}{2rB^2} -l^{-2}&=&\frac{l_P F(M)}{r^3} \,, \\
 \frac{A'}{2rA B} -l^{-2}&=&\frac{l_P F(M)}{r^3} \,, \\
 \frac{2ABA''-AA'B'-BA'^2}{4A^2B^2} - l^{-2}&=&-\frac{2 l_P F(M)}{r^3} \,, 
\end{eqnarray}
where prime denotes $d/d r$. These equations are the same as those encountered in the nonperturbative case discussed in \cite{MZ}. In fact, the $r$-dependence of $T_{\mu\nu}$ for the nonperturbative solution of \cite{MZ} is the same as that of $\left\langle T_{\mu 
 \nu} \right\rangle$. The solution for $A$ and $B$ depend on two integration constants. One of these constants can be set equal to 1 by a time reparametrization, while the other ($C$) can be seen to correspond to the ADM mass of the perturbed solution.  The modified metric is, to first order in  $\hbar$,
\begin{equation} \label{mod21}
ds^2= -\left(\frac{r^2}{l^{2}}-C-\frac{2l_P F(M)}{r}\right) dt^2 +
\left(\frac{r^2}{l^{2}}-C-\frac{2l_P F(M)}{r}\right)^{-1} dr^2 + r^2
d \theta^2 \,.
\end{equation}

The $O(\hbar r^{-1})$ corrections to the metric do not change the value of the ADM mass and therefore this integration constant must be  identified with ADM mass of the unperturbed black hole, $C = M$. For $M>>1$ the $F(M)\rightarrow 0$ and the BTZ black hole is recovered.  Naturally, the metric (\ref{mod21}) has the same form as the conformally dressed black hole metric found in \cite{MZ}. 

The horizons are the real positive roots of \begin{equation}  \label{eqhorm}
r l^2 g^{rr}=r^3-Ml^{2}r-2l_P F(M)l^{2}=0 \,.
\end{equation}
For periodic fields, $F(M,\delta=0)\equiv F^+>0$, and this equation has only one real positive root which, to first order in $\hbar$, is
\begin{equation}  \label{horm}
r_+= \sqrt{M} l + \frac{l_P F^+(M)}{M} \,,
\end{equation}
that is, the horizon of the perturbed solution lies outside the horizon of the unperturbed black hole. In the case of twisted boundary conditions ($\delta \neq 0$), the real part of $F$ can become negative. In particular, for $\delta=\pi$, $F(M,\delta =\pi)\equiv -F^-$ is negative and Eq. (\ref{eqhorm}) has {\em two} real positive roots which, to first order in $\hbar$, are \begin{equation}\label{r+-}
r_+= \sqrt{M} l-\frac{l_P F^{-}(M)}{M} \,,\qquad r_-= \frac{2l_P
F^{-}(M)}{M}\,.
\end{equation}
Thus, an inner horizon results from the antiperiodic boundary conditions. Also, extreme black holes can exist if (\ref{eqhorm}) has a double root. This occurs for 
\begin{equation}\label{extreme}
M= 3 \left( \frac{l_p}{l} F^-(M)\right)^{2/3}\,.
\end{equation}
This relation has a unique solution for $M$ and the horizon radius satisfies 
\begin{equation}\label{r-extreme}
\frac{r_{\rm extreme}}{\lambda_{\rm Compton}}= \frac{3F^-}{16\pi} \simeq O(10^{-3}).
\end{equation}

\subsection{Thermodynamics}

The $O(\hbar)$ corrections to the temperature can be computed using the standard formula (for the nonextreme case) valid if $g_{tt}g_{rr}=-1$: $k_{\rm B}T=-\hbar (g_{tt}')_{r_+} /4\pi$
\begin{equation}
k_{\rm B}T=\hbar \left[ \frac{\sqrt{M}}{2\pi l}\pm\frac{ l_P F^{\pm}(M)}{\pi M
l^{2}}\right]\,.
\end{equation}
The leading term of above expression is the temperature of a BTZ  black hole of mass $M$.  The entropy of the perturbed black hole can be found from the first law of thermodynamics,
$$
dS=\frac{dM}{T}
$$
Integrating this relation, we obtain to first order in $\hbar$,
\begin{equation}\label{S}
S^{\pm}=  \displaystyle\frac{\pi k_{\rm B}}{2l_P}\left[ \sqrt{8GM}l \mp 
l_P\int^{8GM}\frac{F^{\pm}(\zeta)}{\zeta^2}d\zeta\right]+S_0\,.
\end{equation}
The first term is the entropy of a BTZ black hole of mass $M$ and equals  $\frac{1}{4} \times$[Horizon Area] in Planck units\footnote{Here the constant $G$, which was previously set equal to 1/8, has been restored.}. The constant $S_0$ should be chosen so that $S$ vanishes as $r_+ \rightarrow 0$. However, the lower limit of integration in (\ref{S}) cannot be taken equal to zero because the expression for $r_+$ to first order in $l_P$ given (\ref{horm}), (\ref{r+-}) is valid for $8GM>> \left( \frac{l_p}{l} \right)^{2/3}$.

\section{Discussion}

We have computed the exact, circularly symmetric metric produced by the one-loop effective energy momentum tensor of a quantum scalar field conformally coupled to gravity. The solution is a modification of the BTZ solution with $O(\hbar r^{-1})$-corrections. These corrections produce a curvature singularity at the origin, so the perturbed geometry is not a constant curvature spacetime as the unperturbed black hole. 

The metric (\ref{mod21}) has the same form as its nonperturbative classical counterpart \cite{MZ}.  It seems a remarkable coincidence that the perturbative solution around $\Psi=0$ should obey the same equations [c.f., Eqs. (14-16)], as the one found in the fully nonperturbative field configuration, $\Psi = \sqrt{ \frac{8r_+}{ \pi(2r+r_+)}}$ (see \cite{MZ}). This can be understood as a necessary consequence of the conformal coupling and of the pecularities of 2+1 gravity. 

Indeed, Einstein's equations in the gauge  $g_{tt} g_{rr} = -1$, imply that the static, spherically symmetric metric is totally determined, up to two constants of integration, whenever the condition $T^{\mu}_{\mu}=0$ holds.  Now, the absence of trace anomalies in 2+1 dimensions \cite{birrell} guarantees that the tracelessness of $T^{\mu}_{\nu}$, due to the conformal coupling, remains true in the semiclassical approximation. Therefore, both the fully nonperturbative Einstein equations and the perturbative semiclassical ones have the same solutions, except that the constants of integration are fixed in the first case and not in the latter.

A related issue is the fact that quantum gravity in 2+1 dimensions is renormalizable and finite \cite{witten}. This means that the only radiative corrections to the geometry are produced by the quantum excitations of the matter fields, since the perturbative expansion receives no corrections from graviton loops.  

We observe that the corrections to the temperature and entropy are linear  in $F(M)$. Since $F(M) \sim e^{-\pi M}$ for large $M$, these corrections are strongly supressed in the massive holes.  Thus, the  back-reaction becomes large for small masses compared to the Planck mass (which, in three dimensions is independent of $\hbar$).  

As seen from Eq. (\ref{r-extreme}), the Schwartzschild radius of an extreme the black hole is of the same order as the Compton wavelength of the state. This implies that extreme holes should be viewed as fully quantum systems, for which the semiclassical approximation could not be trusted.

Another point of interest has to do with length scales. The corrections induced by $\left\langle T_{\mu \nu} \right\rangle$ depend on the ratio $l_P/l$, that is, the size of the Planck
length (which is linear in $\hbar$) relative to the radius of the universe (fixed by the cosmological constant). Therefore, for small cosmological constant ($l >>1$) the perturbation on the geometry produced by radiation fields is negligible.


\begin{acknowledgements}

Many useful discussions with M. Ba\~{n}ados, M. Contreras, N. Cruz, J. Gamboa, A. Gomberoff and R. Troncoso are greatly acknowledged. This work was supported in part by grants 1940203, 2940012 and 1960229 from FONDECYT (Chile), and 049531ZI (USACH). The
institutional support of a group of Chilean companies (EMPRESAS CMPC, CGE, CODELCO, COPEC, MINERA ESCONDIDA, NOVAGAS, BUSINESS DESIGN ASS. and XEROX-CHILE) is also recognized.

\end{acknowledgements}


\begin{references}

\bibitem{hawking} S. W. Hawking,  Comm. Math. Phys. {\bf 43}, 199 (1975). 

\bibitem{birrell}N. D. Birrell and P. C. W. Davies, {\em Quantum Fields in
Curved Spaces} (Cambridge University Press, 1982).

\bibitem{york}. J. W. York, Jr., Phys. Rev. {\bf31}, 775 (1985).

\bibitem{witten}E. Witten, { Nucl. Phys.} {\bf B 311}, 46 (1988).

\bibitem{BTZ} M. Ba\~{n}ados, C. Teitelboim and J. Zanelli, Phys. Rev. Lett.
{\bf 69}, 1849 (1992).

\bibitem{BHTZ} M. Ba\~{n}ados, M. Henneaux, C. Teitelboim and J.
Zanelli, { Phys. Rev. D} {\bf 48}, 1506 (1993).

\bibitem{carlip1}  S.~Carlip, {``Lectures on (2+1)-Dimensional Gravity",} Davis
preprint UCD-95-6, gr-qc/9503024 (1995).	

\bibitem{carlip2} S.~Carlip, {Class. Quantum Grav.}~{\bf 12},
2853 (1995). 

\bibitem{LO} G. Lifschytz and M. Ortiz, Phys. Rev. D {\bf 49}, 1929 (1994).

\bibitem{SM1} K. Shiraishi and T. Maki, Phys. Rev. D {\bf 49}, 5286 (1994).

\bibitem{MZ} C. Mart\'{\i}nez and J. Zanelli, Phys.Rev. D {\bf 54}, 3830
(1996).	 

\bibitem{AIS} S. J. Avis, C. J. Isham and D. Storey, Phys. Rev. D {\bf 18},
3565 (1978).

\bibitem{steif} A. R. Steif, Phys. Rev. D {\bf 49}, 585 (1994).


\end{references}
\end{document}